\documentclass[aps,pre,showkeys,onecolumn,superscriptaddress,notitlepage,
floats,floatfix,longbibliography]{revtex4-1}
\usepackage{amssymb,amsmath}
\usepackage{graphics,rotating}
\usepackage[dvipsnames]{xcolor}
\usepackage{wasysym}
\def \mm  {{\bm m}}

\def \Cmm {C_{\mu\mu}}

\def \EP   {E_{\rm p}}
\def \Bh   {\hat{B}}
\def \mB   {\mathcal{B}}

\def \Pit {\Pi_{\rm }}
\def \Pip {\Pi_{\rm p}}
\def \Pif {\Pi_{\rm f}}
\def \mDp {\mathcal{D}_{\rm p}}
\def \lamq {\lambda_{\rm q}}

\def \epsl  {\varepsilon_{\ell}}
\def \epsilonf {\varepsilon_{\rm f}}
\def \epsilonp {\varepsilon_{\rm p}}
\def \mF {\mathcal{F}_{\rm inj}}
\def \mD {\mathcal{D}_{\rm f}}
\def \mP {\mathcal{P}}
\def \FF     {\bm{F}}

\def \Sthree {S_{\rm 3}}
\def \Stwo {S_{\rm 2}}
\def \deltau {\delta u}
\def \Sp    {S_{\rm q}}
\def \Ell   {\bm{\ell}}
\def \pp   {{\rm q}}
\def \einj {\varepsilon_{\rm inj}}
\def \nuf  {\nu_{\rm f}}
\def \tauL {\tau_{\rm L}}

\def \Lint {L_{\rm }}
\def \kI {k_{\rm L}}
\def \kd {k_\eta}
\def \keta {k_{\eta}}
\def \uhat {\hat{\bm{u}}}
\def \mS   {\mathcal{S}}
\def \mup  {\mu_{\rm p}}
\def \kmax {k_{\rm max}}
\def \tN   {{\tt N}}
\def \ua   {u_{\alpha}}
\def \ub   {u_{\beta}}
\def \ug   {u_{\gamma}}
\def \xa   {x_{\alpha}}
\def \xb   {x_{\beta}}
\def \Sab  {S_{\alpha\beta}}
\def \ab   {\alpha\beta}
\def \muf  {\mu_{\rm f}}
\def \rhof {rho_{\rm f}}
\def \mC  {\mathcal{C}}
\def \Cab {C_{\alpha\beta}}
\def \Kp  {k_{\rm p}}

\def \uhat {\hat{u}}
\def \Lf {L_{\rm inj}}
\def \LK {L_{\rm K}}
\def \Lp {L_{\rm p}}
\def \Stwo {S_{\rm 2}}

\def \Tr  {{\rm Tr}}

\def \mE {\mathcal{E}}
\def \mEf {\mathcal{E}_{\rm f}}
\def \mEp {\mathcal{E}_{\rm p}}
\def \mL {\mathcal{L}}

\def \mL   {\mathcal{L}}

\def \mP {\mathcal{P}}

\def \rhof {\rho_{\rm f}}

\def \uu  {{\bm u}}
\def \kk  {{\bm k}}

\def \teta {\tau_{\rm \eta}}
\def  \xx  {{\bm x}}

\def  \ua  {u_{\alpha}}

\def  \ub  {u_{\beta}}

\def   \kstar {k_{\ast}}

\def \taup {\tau_{\rm p}}

\newcommand{\bra}[1]{\left\langle #1\right\rangle}

\def \De  {\mbox{De}}
\def \Wi  {\mbox{Wi}}
\def \Rey  {\mbox{Re}}
\def \Rel  {\mbox{Re}_\lambda}

\def \kf  {k_{\rm inj}}

\def \urms  {u_{\rm rms}}

\def \zetap {\zeta_{\rm q}}
\def \zetaq {\zeta_{\rm q}}
\def \zetat {\zeta_{\rm 2}}

\def \deltaq {\delta_{\rm q}}

\newcommand{\dela}[1]{\frac{\partial #1}{\partial x_{\alpha}} }

\newcommand{\delg}[1]{\frac{\partial #1}{\partial x_{\gamma}} }

\newcommand{\eq}[1]{~(\ref{#1})}

\newcommand{\Fig}[1]{Fig.~(\ref{#1})}
\newcommand{\subfig}[2]{Fig.~(\ref{#1}#2)}

\newcommand{\bfig}{\begin{figure}}
\newcommand{\efig}{\end{figure}}
\newcommand{\bc}{\begin{center}}
\newcommand{\ec}{\end{center}}
\newcommand{\bea}{\begin{eqnarray}}
\newcommand{\eea}{\end{eqnarray}}

\begin{document} 
\title{Large is different: non-monotonic behaviour of elastic range scaling
  in polymeric turbulence at large Reynolds and Deborah numbers}
\author{Marco E. Rosti}
\email{marco.rosti@oist.jp}
\affiliation{Complex Fluids and Flows Unit, Okinawa Institute of Science and
  Technology Graduate University, 1919-1 Tancha, Onna-son, Okinawa 904-0495, Japan}
\author{Prasad Perlekar}
\affiliation{TIFR Centre for Interdisciplinary Sciences,
  Tata Institute of Fundamental Research,  Gopanpally, Hyderabad 500046,
  India.}
\author{Dhrubaditya Mitra}
\affiliation{ Nordita, KTH Royal Institute of Technology and
Stockholm University, Roslagstullsbacken 23, 10691 Stockholm, Sweden}

\begin{abstract}
We use direct numerical simulations
  to study homogeneous, and isotropic turbulent flows
  of dilute polymer solutions at high Reynolds and Deborah numbers.
  We find that for small wavenumbers $k$, the kinetic energy spectrum shows
  Kolmogorov--like behavior which crosses over at a larger $k$ to a novel,
  elastic scaling regime, $E(k) \sim k^{-\xi}$, with $\xi \approx 2.3$.
  We study the
  contribution of the polymers to the flux of kinetic energy through scales, and find that
  it can be decomposed into two parts: one increase in
  effective viscous dissipation, and a purely 
  elastic contribution that dominates over the nonlinear flux
  in the range of $k$ over which the elastic scaling is observed.
  The multiscale balance between the two fluxes determines the crossover
  wavenumber
  which depends non-monotically on the Deborah number.
  Consistently, structure functions also show two
  scaling
  ranges, with intermittency present in both of them in equal measure.
\end{abstract}

\maketitle
\section{Introduction}
\label{sec:intro}
Since the discovery of turbulent drag reduction by Toms \cite{toms49},
turbulent flows with small amount of long-chained polymers have remained
an exciting field of research.
In addition to polymer concentration,
two dimensionless numbers, the  Reynolds number and the
Deborah number, are necessary to describe such a turbulent flow. 
The former estimates the importance of the inertial term in the
Navier--Stokes equation compared to the viscous term, and the latter
is the ratio of the characteristic time scale of the polymers over the
typical time scale of the large scale eddies in the  turbulent flow.
The turbulent drag reduction appears at both large Reynolds and Deborah numbers.
Evidently, it is not possible to study drag reduction in homogeneous
and isotropic turbulent flows, nevertheless such flows are
studied, since the pioneering work by Tabor and De Gennes \cite{tab86},
in search of deeper insights.

The elementary effect of the addition of polymers to a fluid is
an increase in the effective viscosity of the solution \cite{hin77,lum73}.
Nevertheless, there can be net reduction of the dissipation of kinetic
energy \cite{bon93,van1999decay,kal_poly04,per+mit+pan06,Per09,per+mit+pan10,
cai2010dns}
because the presence of polymers changes the turbulent cascade \textit{qualitatively}. 
Significant
theoretical~\cite{lum73,tab86,bhattacharjee1991drag,thirumalai1996polymer,fux03},
numerical~\cite{vai03,ben03,de2005homogeneous, berti2006small, per+mit+pan06,
  pet07,per+mit+pan10, cai2010dns, de2012control, watanabe2013hybrid, fathali2019spectral,
 watanabe2014power,nguyen2016small, valente2014effect,valente2016energy}, and
experimental~\cite{fri70,mcc77,bon93,bon05,lib06,oue09,vonlanthen2013grid,
  zhang2021experimental} efforts
have gone into elucidating the nature of the turbulent
energy cascade in the presence of polymers.
It is now reasonably well established \cite{per+mit+pan06,per+mit+pan10} that
for large enough scale separation between the energy injection
scale, $\Lf$, and the Kolmogorov scale, $\LK$,
there exists an intermediate scale $\Lp$ such that
for scales $\Lp < r < \Lf$ the energy cascade is practically
the same as that of a Newtonian flow, with the
the second order structure function
$\Stwo(r) \sim r^{2/3}$ and the 
shell-integrated energy spectrum being
$E(k) \sim k^{-5/3}$.
For scales $r$ in the range  $\LK < r < \Lp$,
energy is transferred from the fluid to the polymers
and the kinetic energy spectrum is steeper than the Kolmogorov spectrum
or in other words the second order structure function
increases faster with $r$ than $r^{2/3}$.
Using the concept of scale-dependent Reynolds
number \cite{LLfluid}, we may identify the 
flow at scale $r < \Lp$ (also valid for
$r < \LK$ ) with elastic turbulence --
random viscoelastic smooth flows at very small Reynolds number.
The spectrum for elastic turbulence  is expected to
be $E(k) \sim k^{-\xi}$ with
$\xi > 3$ \cite{fux03,gro+ste00,steinberg2021elastic}.
Is there a new scaling range
for $r < \Lp$ over which $\Stwo(r) \sim r^{\zetat}$
with $2/3 < \zetat < 2 $ ?
This question could not be probed with the low-Reynolds and low-Deborah
simulations quoted above. 
Recent experiments \cite{zhang2021experimental}
had tentatively suggested that a new scaling range indeed appears,
although the evidence is not yet unequivocal. 
Experiments \cite{xi2013elastic,zhang2021experimental} also showed that,
contrary to Lumley's arguments \cite{lumley1969drag},
the scale $\Lp$ does depends on the concentration of polymers.

Here we present evidence, from the highest resolution 3D simulations
of polymeric fluids, that indeed there is a range of scales
$r$ over which the structure function $\Stwo(r)$ seems to
show scaling consistent with recent experimental
results \cite{zhang2021experimental}.
We also show that the new scaling is a purely elastic effect,
and that this elastic behaviour is non-monotonic
in the Deborah number.

\section{Results}
\subsection{Governing equations}
We use direct numerical simulations to study three dimensional homogeneous isotropic turbulence with
polymers \cite{pet66,war72,arm74,hin77,Bir87,Pha02}.
These are represented by a second rank tensor, $\mC$ with components $\Cab$, which
emerges as the thermal average of the tensor-product of the polymer end-to-end
distance with itself.
The polymer molecules are assumed to have a single relaxation time $\taup$.
The dynamical equations are:
\begin{subequations}
\label{eq:all}
  \begin{align}
\label{eq:ns}
\rhof \left(\frac{\partial \ua}{\partial t}
+ \frac{\partial \ua\ub}{\partial \xb} \right) &=
- \frac{\partial p}{\partial \xa} +
\frac{\partial}{\partial \xb}\left( 2 \muf S_{\ab} 
+  \frac{\mup}{\taup} f \Cab\right) +  F_{\alpha}\/, \\
\label{eq:poly}
\frac{\partial \Cab}{\partial{t}} + u_\gamma \frac{\partial C_{\alpha \beta}}{\partial x_\gamma} &=
C_{\alpha\gamma}\delg{\ub} +C_{\gamma\beta}\dela{\ug}
-\frac{f\Cab - \delta_{\ab}}{\taup}\/, \\
\frac{\partial \ua}{\partial \xa} &= 0\/.
\end{align}
\end{subequations}
Here $\uu$ is the velocity, $\rhof=1$
and $\muf$ are the density and dynamic viscosity of the fluid,
$p$ is the pressure,
$\mup$ is the polymer viscosity,
and $\mS$ is the rate-of-strain tensor with components
$S_{\ab}$ defined as
$S_{\ab}=\left( \partial \ua/ \partial \xb + \partial \ub / \partial \xa\right)/2$.
The function $f$ is equal to unity ($f=1$) in the purely elastic Oldroyd-B model, and to $f = \left(\mL^2-3 \right)/\left(\mL^2-C_{\gamma\gamma}\right)$ in the FENE-P
model~(where $\mL$ is the maximum allowed extension of the polymers) exhibiting both
shear-thinning and elasticity.
The polymer timescale is the relaxation time $\taup$ and its concentration is related to
the value of $1+\mup/\muf$; the value chosen in this work corresponds, roughly, to
$100$ ppm for polyethylene oxide \cite{virk_1975a}.
Note that, we work in the dilute limit where polymer concentration is assumed to be
homogeneous.
Turbulence is sustained by the external force in the momentum equation, $\FF$; we
use the spectral scheme from \citet{eswaran1988forcing} to randomly inject
energy to the low-wavenumber shells with $\kf = \left( 1\le k \le2 \right)$.
Note that, the scaling behavior in wavenumbers much larger than $\kf$  are
    independent of the choice of $\kf$.
In the statistically stationary state of turbulence, the injected energy is dissipated
by both the fluid ($\epsilonf$) and the polymers ($\epsilonp$), thus
$\einj = \epsilonf + \epsilonp$, where
\begin{equation}
  \epsilonf = \frac{2 \muf}{\rhof}\bra{S_{\ab}S_{\ab}}\/,\quad\/
  \epsilonp = \frac{\mup}{2 \rhof \taup^2}\bra{f \left( f \Cmm - 3 \right)}\/.
\label{eq:eps}
\end{equation}

To compare, we also solve for the Navier--Stokes equations without any polymer
additive -- we call this the Newtonian simulation.  

\subsection{Theoretical background}
Let us briefly recall essential features of fluid turbulence without polymeric
additives \cite{Fri96}.
The flow is determined by one dimensionless number,
$\Rey=\urms/(\kf\nuf)$, where $\urms$ is the root-mean-square velocity and
$\nuf = \muf/\rhof$ is the kinematic viscosity of the fluid.
Turbulent flows possess a range of length scales and corresponding time scales.
The statistical properties of such flows are characterized by the scaling exponents,
$\zetap$ of the $\pp$-th order longitudinal structure functions, $\Sp$,
defined by:
\begin{subequations}
  \label{eq:Sp}
  \begin{align}
    \Sp(\ell) &= \bra{\deltau(\ell)^q} \sim \ell^{\zetap} \quad \text{where}
        \label{eq:zetap}\\
    \deltau(\ell) &\equiv \left[\uu(\xx+\Ell) - \uu(\xx)\right]
    \cdot\left(\frac{\Ell}{\ell}\right) \/.
  \end{align}
\end{subequations}
Here $\bra{\cdot}$ denotes averaging over the statistically stationary state of turbulence.
The $\pp$-th order structure function is the $\pp$-th order moment of the probability
distribution function of velocity difference across a length scale $\ell$.
The scaling behavior of the structure function, \eq{eq:zetap}, holds for
$ \eta \ll \ell \ll \Lint$ where $\eta \equiv (\nuf^3/\einj)^{1/4}$ is
called the viscous scale and $\Lint$ is called the integral scale.
In practice, $\Lint = 2\pi/\kI$ is of the same order of $\Lf = 2\pi/\kf$ and we will
use them interchangeably.

The shell-integrated energy spectrum in Fourier space 
\begin{equation}
  E(k) \equiv \int_{\lvert\mm\rvert=k}d^3\mm \bra{\uhat(\mm)\uhat(-\mm)}\/,
  \label{eq:Ek}
\end{equation}
where $\uhat(\mm)$ is the Fourier transform of the velocity field $\uu(\xx)$,
is itself the Fourier transform of the second order structure function $\Stwo(\ell)$.
The theory of Kolmogorov gives $\zetap = q/3$ and consequently $E(k) \sim k^{-5/3}$,
when $k$ lies within the \textit{inertial range},
$\kf \ll k \ll \kd$, with $\kd \sim 1/\eta$.
The turbulent velocity fluctuations are non-Gaussian in two ways.
First, the odd-order structure functions are non-zero, in particular the third order
structure function satisfies the most celebrated exact relation in turbulence, i.e.~the
four-fifth law, $\Sthree(\ell) = -(4/5)\einj \ell$ -- this result is the cornerstone of
Kolmogorv's theory of turbulence.
Second, the scaling exponents $\zetap$ are a nonlinear convex function of $\pp$
-- a phenomena called \textit{intermittency}.

{In the presence of polymers, we, in addition, consider 
\begin{equation}
  \EP(k) \equiv \left(\frac{\mup}{\rhof \taup}\right)\int_{\lvert\mm\rvert=k}d^3\mm\bra{
    \Bh_{\gamma\beta}({\mm})\Bh_{\beta\gamma}(-\mm)}\/,
\label{eq:Ep}
\end{equation}
where the matrix $\mB$ with components, $B_{\alpha\gamma}$, is the (unique) positive symmetric
square root of the matrix $\mC$, i.e.,
$\Cab = B_{\alpha\gamma}B_{\gamma\beta}$ \cite{balci2011symmetric,nguyen2016small}. 
For the Oldroyd-B model the total energy in the polymeric mode is given by  
\begin{equation}
	\mE_{\rm p} \equiv \frac{1}{2}\int dk \EP(k) = \frac{\mup}{2\rhof\taup}\bra{C_{\mu\mu}}.
	\label{eq:mEp}
\end{equation}
}
\begin{figure*}
\centering
\includegraphics[width=0.85\textwidth]{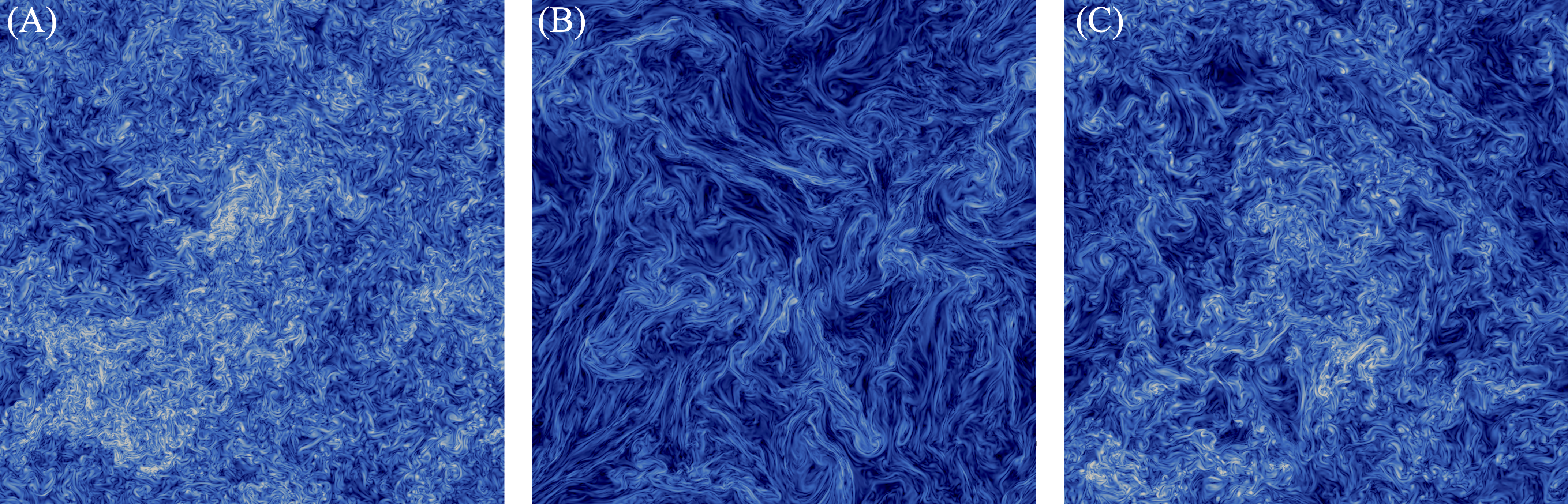}
\caption{\textbf{Instantaneous snapshots of the turbulent flows.} (A) Newtonian
  and viscoelastic fluid with (B) $De\approx 1$ and (C) $De\approx 25$, at a nominal microscale Reynolds number $\Rel \approx 400$ for the Newtonian case ((A): $\Rel = 390$, (B) $\Rel \approx 740$, (C) $\Rel \approx 447$). The color contour shows the magnitude of the
  vorticity field, with the color-scale going from $0$ (blue) to the
  maximum (white). The figures are two-dimensional cuts of the three-dimensional periodic cube passing trough the middle of the domain.}
\label{fig:cover}
\end{figure*}
The presence of polymers introduces also a new dimensionless number which is the
ratio of the polymeric time scale $\taup$ over a characteristic time scale of
the flow.
As the turbulent flow has many time scales, it is common to define the
Deborah number
$\De \equiv \taup/\tauL$, where $\tauL = \Lint/\urms$ is the
large-eddy-turnover-time,
and the Weissenberg number $\Wi \equiv \taup/\teta$, where
$\teta = \eta^2/\nu$ \cite{ben18}.

In \Fig{fig:cover} we show typical pseudocolor plots
of vorticity from Newtonian and viscoelastic simulations.
The flow is qualitatively strongly affected by the presence of polymers, and
small-scale vorticity structures are smoothened by the presence of the polymers,
as can be seen by comparing \subfig{fig:cover}{A} and \subfig{fig:cover}{B},
see also Ref.~\cite{de2005homogeneous,per+mit+pan06,per+mit+pan10}. 
Surprisingly, as the Deborah number is increased beyond unity, this 
qualitative trend is reversed, compare \subfig{fig:cover}{B} and
\subfig{fig:cover}{C}. 
In \subfig{fig:cover}{C} small scale structures in vorticity
reappears but at the same time we still find 
elongated structures although their length scales
are smaller than their counterparts in 
\subfig{fig:cover}{B}.

\begin{figure*}
\centering
\includegraphics[width=0.95\textwidth]{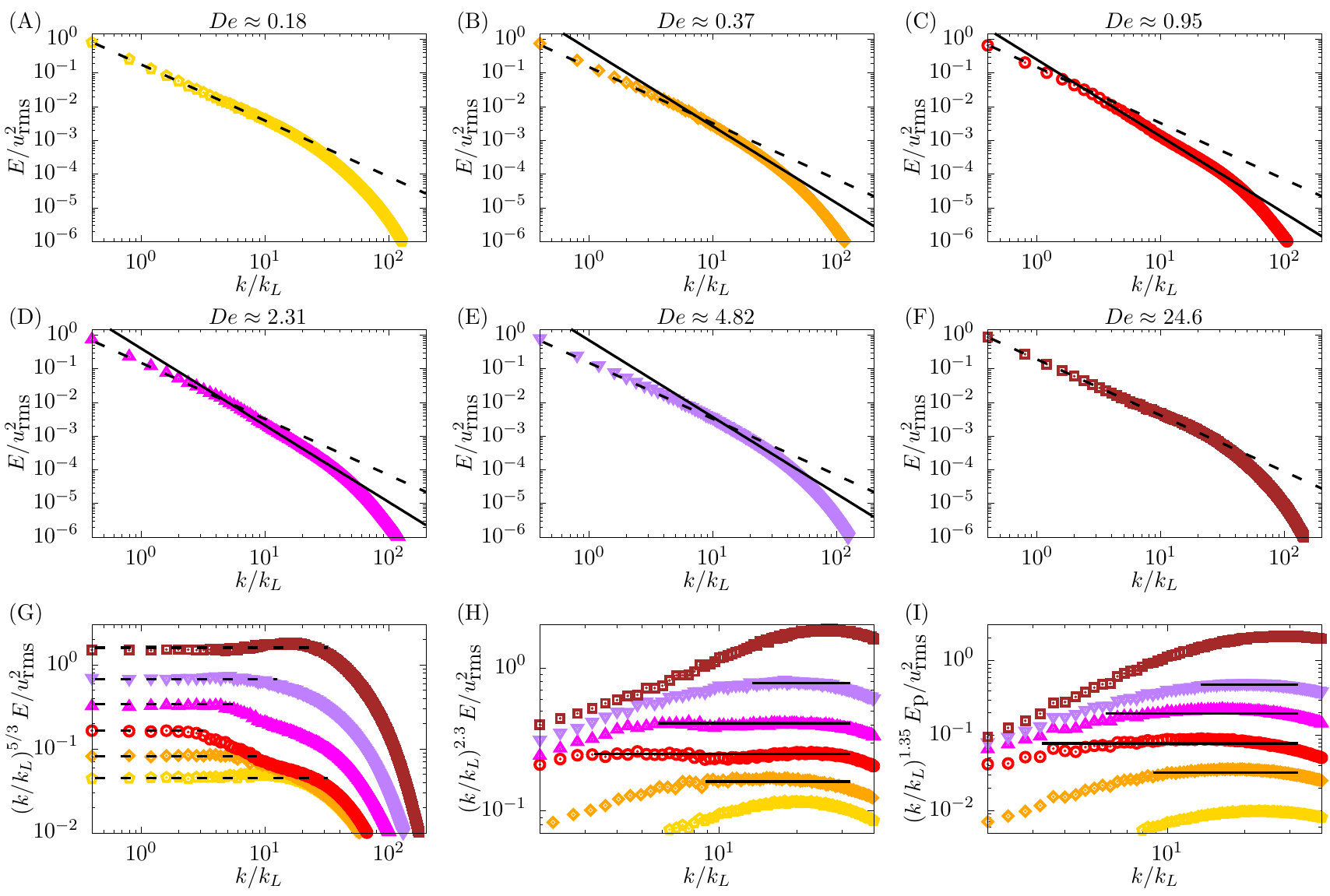}
\caption{\textbf{Kinetic and polymer energy spectra.} (A) to (F): Energy spectra for different Deborah numbers showing how the elastic range changes with $\De$. (G) and (H): Compensated kinetic energy spectra showing the emergence of two scaling regions, Kolmogorov ($-5/3$) and elastic ($-2.3$) scaling respectively. In all the previous panels, the dashed and solid lines represent the $-5/3$ and $-2.3$ scalings. (I): Compensated polymer energy spectra for different Deborah numbers. The solid line represent the scaling $k^{-\psi}$ with $\psi\approx 1.35$. The scaling laws in (G--I) extend over the scales found from the cross overs extracted from \Fig{fig:flux}. The abscissa are normalized with the integral length scale wavenumber $k_L$.}
\label{fig:spectra}
\end{figure*}
\subsection{Kinetic and polymer energy spectra}
In \Fig{fig:spectra} we plot the turbulent kinetic energy $E(k)$ for several values of
Deborah number $\De$.
For the small Deborah numbers, e.g.~$\De \approx 0.18$, we observe, practically,
the same behavior as Kolmogorov turbulence, with $E(k) \sim k^{-5/3}$ for the
inertial range.
As the Deborah number increases the range over which the Kolmogorov scaling is valid
shrinks to smaller $k$, and at intermediate $k$ a new range over which
$E(k) \sim k^{-\xi}$ with $\xi \approx 2.3$ emerges.
We call this new scaling range the \textit{elastic range}.
The spectra, in general, has three characteristic length scales
(or equivalently wavenumbers).
The largest is the one where energy is injected by stirring, the integral scale,
$\Lint \approx \Lf$.
Next is the scale at which the Kolmogorov scaling crosses over to elastic scaling,
$\Lp$ (corresponding wavenumber $\Kp = 2\pi/\Lp$).
Last is the scale at which elastic scaling crosses over to the dissipative range,
which we call the dissipative scale $\eta$ (corresponding wavenumber $\keta = 1/\eta$).
The Kolmogorov scaling is observed over the range $ \kf < k < \Kp$ and the elastic range
over the wavenumber range $\Kp < k <\keta$.
The elastic range spans over the maximum range of wavenumber for
$\De \approx 1 $; as $\De$ is increased further the elastic range
begins to shrink and the Kolmogorov range begins to grow again.
Eventually, the elastic range practically disappears at $\De \approx 5$ --
the classical Kolmogorov range is restored.
This remarkable behavior is better elucidated by plotting the two compensated
spectra $k^{5/3}E(k)$ and $k^{\xi}E(k)$ in \subfig{fig:spectra}{G} and
\subfig{fig:spectra}{H}, respectively.
In other words, our results clearly show that the wavenumber $\Kp$ depends
non-monotonically on the Deborah number, being maximum for $\De \approx 1$.
The non-monotonic behaviour of the polymeric flow for $\De \approx 1$ can be
also qualitatively appreciated by observing \Fig{fig:cover}.

Although, the steepening of the spectra beyond a certain wavenumber $k=\Kp$
have been observed before in both direct numerical
simulations~\cite{de2005homogeneous,per+mit+pan06,berti2006small,
  per+mit+pan10,valente2016energy} and experiments~\cite{vonlanthen2013grid},
this was mostly confined to the dissipation range due to the small separation of
scales related to the Reynolds number considered; only recent
experiments~\cite{zhang2021experimental}, demonstrated for the first time the
emergence of this elastic scaling.
Ref.~\cite{zhang2021experimental} also found that $\Lp$ increases with the
polymer concentration but they did not investigate how it behaves as a function
of the Deborah number.
Also, we find for the first time both Kolmogorov and elastic scaling simultaneously
valid for different ranges of wavenumbers by virtue of running the largest
simulation of polymeric turbulence so far and we also uncover the
non-monotonic behavior of $\Lp$ as a function of the Deborah number.
We have also confirmed that these results are robust with respect to change
  in spatial and temporal resolutions, see Fig.~(S2).

The previous results have been obtained for the purely elastic Oldroyd-B model.
Before we explore further the elastic scaling, it is worth mentioning that we
performed the simulation for $\De \approx 0.95$  with two additional models of
polymeric fluids -- the inelastic, shear-thinning Carreau Yasuda model and the
FENE-P model, which models both the elastic and shear-thinning behaviour of
polymeric fluids.
We find that the new scaling at intermediate scales is a purely
elastic effect, which completely disappear in the absence of elasticity,
while it is
reduced when shear-thinning is present together with elasticity
(see Fig.(S3) in the Supplementary Materials).
We have also observed that if the parameter  $\mL$ (the maximum
possible extension of the polymers) of the FENE-P model is varied within
a reansonable range the elastic scaling remains practically unchanged.
For too small a value of $\mL$ the elastic scaling range can
disappear, see Fig.~(S3B). 

\begin{figure*}
\centering
\includegraphics[width=0.95\textwidth]{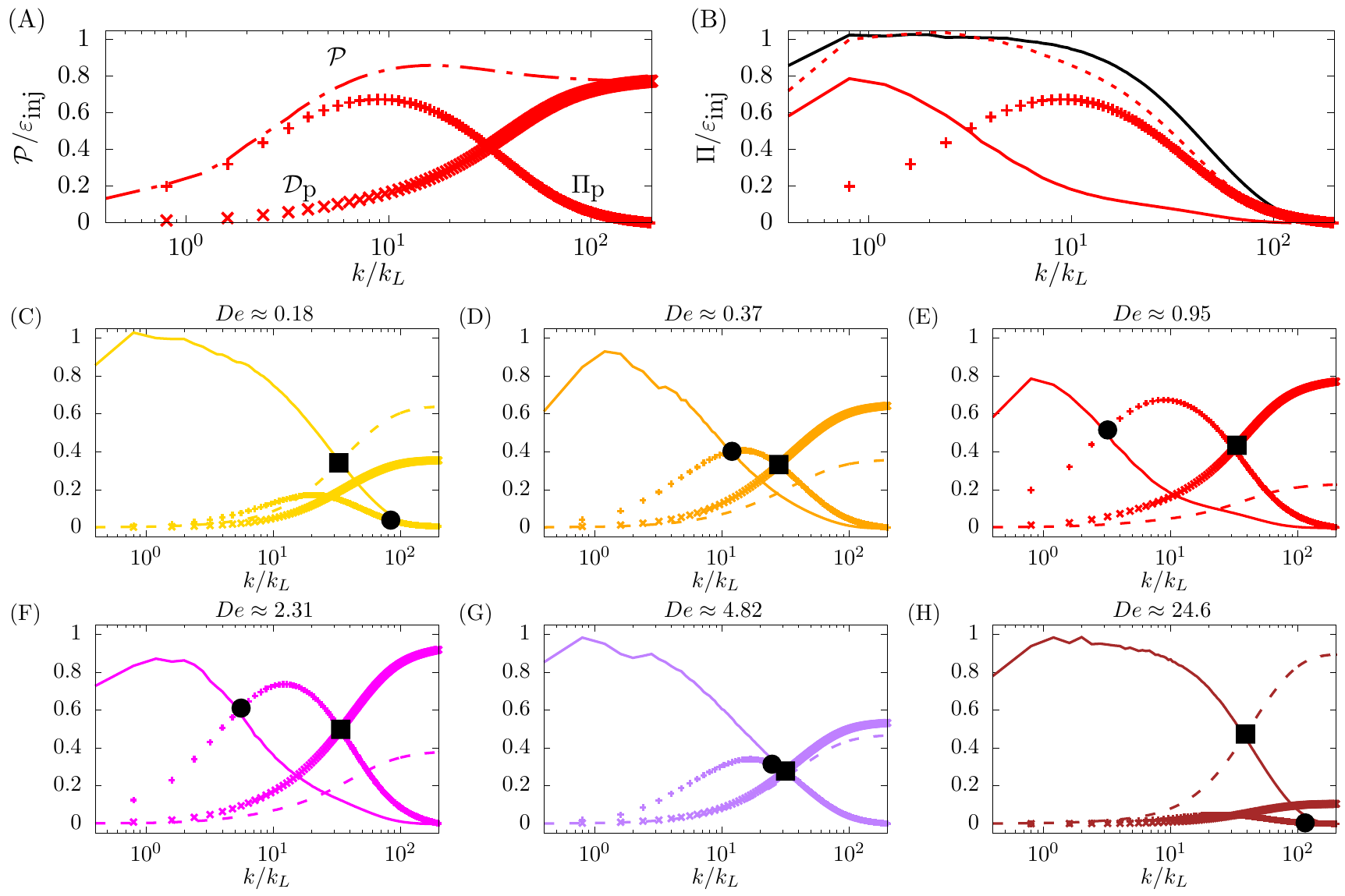}
\caption{\textbf{Scale by scale energy budget} (A): The polymer contribution to the spectral energy balance, $\mP$, (dash-dotted line) is decomposed into a ($\times$) pure polymer dissipation term, $\mDp$ and a ($+$) pure polymer energy flux, $\Pip$, see \eq{eq:pflux}. (B): The sum of the (solid line) non-linear energy flux $\Pif$ and of the ($+$) polymer flux $\Pip$ provide a (dotted line) total flux $\Pit = \Pif+\Pip$ extending over a range comparable to the Newtonian case (black line). (C) to (H): The panels show the non-linear energy flux $\Pif$ (solid line), fluid dissipation $\mD$ (dashed line),  polymer flux, $\Pip$ ($+$), and polymer dissipation $\mDp$ ($\times$) for different Deborah numbers. The filled circles represent $\Kp$ and the filled squares $\keta$, used in \subfig{fig:spectra}{G--I} as the extension of the scaling laws. As $De$ approaches unity, the polymer flux and dissipation grow, while they decrease for larger values. The abscissa are normalized with the integral length scale wavenumber $k_L$. {The color scheme is the same used in \Fig{fig:spectra}.}}
\label{fig:flux}
\end{figure*}
\subsection{Scale-by-scale energy budget}
In turbulence, to understand the energy spectra we have to study the flux of energy
through scales \cite{Fri96,pope_2001a,Verma2019energy}.
For polymeric turbulence, the flux in Fourier space have been studied before
by Refs.~\cite{valente2014effect,valente2016energy} and their real space analog
in Ref.~\cite{de2005homogeneous}.

{To obtain the flux of kinetic energy in Fourier space, transform \eq{eq:ns} to
Fourier space, multiply by $\uhat(-\kk)$, integrate over the solid angle
$d\Omega$ and over $k$ from $0$ to $K$, and average over the statistically
stationary state of turbulence to obtain
\begin{equation}
     \einj = \epsilonf + \epsilonp = \Pif(K) + \mD(K) + \mP(K) + \mF(K)\/,
\label{eq:eflux}
\end{equation}
where $\Pif$, $\mD$, $\mP$ and $\mF$ are the
contributions from the nonlinear term, the viscous term, the polymeric stress 
and
the external force in \eq{eq:ns} (see the Supplementary Materials for a full derivation).
The first equality of \eq{eq:eflux} follows from statistical stationarity. 
For $K \gg \kf$, the external force is zero hence its contribution
to the flux, $\mF$ is constant, i.e., $\mF(K \gg \kf) = \einj$.
In the absence of polymers, $\mP = 0$, and, since in the inertial range the 
dissipative contribution $\mD$ is negligible,
$\Pif(K \gg \kf) = -\einj $ is a constant.
The Kolmogorov four-fifth law follows from this statement~\cite{Fri96}. 
In addition, if we assume that scaling we obtain 
$E(k) \sim k^{-5/3}$.  
}

The novel physics of this problem is elucidated by studying the contribution
from the polymers, $\mP$.
In \subfig{fig:flux}{A} we show a representative plot of $\mP(K)$ as a function of
$K$, plotted as a dashed-dotted line.
It is well established \cite{Bir87,hin77,lum73} that one of the
effects of addition of polymers to flows is the increase of dissipation,
but $\mP(K)$ is not a purely dissipative term, as shown by its non-monotonicty
with $K$.
{This feature has been modelled before by a wavenumber dependent
  effective viscosity~\cite{per+mit+pan06}.
  Here we try a different approach. 
  We  separate the part of $\mP(K)$ which is
  purely dissipative, $\mDp(K)$,
  such that at large $K$ such a term should have the same
  asymptotic dependence on $K$ as $\mD(K)$.
  We further demand that as $K\to \infty$,
  $\mDp(K) \to \epsilonp$. Hence we obtain:
\begin{subequations}
  \begin{align}
    \mP(K) &= \Pip(K) + \mDp(K)\/,\quad\text{where} \label{eq:pflux} \\
    \mDp(K) &\equiv \frac{\epsilonp}{\epsilonf}\mD(K) \/.\label{eq:pdiss}
  \end{align}
\end{subequations}}
We plot $\Pip$ and $\mDp$ individually in \subfig{fig:flux}{A}.
Remarkably, $\Pip$ has the same qualitative behavior as $\Pif$, the nonlinear flux.
In \subfig{fig:flux}{B} we plot both $\Pip$ and $\Pif$ denoted by the symbol $+$ and a
continuous line respectively.
We find that for small $K$, $\Pif$ is dominant and $\Pip$ is insignificant.
At a certain scale $\kstar$ the two fluxes cross each other.
Beyond $\kstar$, $\Pip$ is the dominant partner and $\Pif$ is negligible.
At very large $K$, well within the dissipation range, both $\Pif$ and $\Pip$
go to zero.
The sum of these two fluxes
$\Pit \equiv \Pip + \Pif$
is practically a constant for all $K \ll \kd$.
In the same figure, \subfig{fig:flux}{B}, we also plot, as a black line, the contribution
to the flux from the nonlinear term for a simulation with no polymers.
Clearly, the flux that is carried by the nonlinear term in the absence of polymers is
carried by both $\Pif$ and $\Pip$ in the presence of polymers:
at small $K$ the flux is carried mainly by $\Pif$ and at large $K$ the flux is
carried {mainly} by $\Pip$.
The crossover between this two fluxes happens at $\kstar$ which we identify with
$\Kp$.
The fluxes clearly illustrate and substantiate what we already observed in the energy
spectra: for $k < \Kp$ the turbulence is Kolmogorov-like whereas for
$\Kp < k < \keta$ the polymeric flux $\Pip$ dominates and is approximately a constant.
We define the range of Fourier modes, $\Kp < k < \keta$ as the \textit{elastic range}
with $\Kp$ precisely defined as $\Pif(\Kp) = \Pip(\Kp)$.

{We emphasize, that the decomposition in \eq{eq:pflux} and \eq{eq:pdiss} 
is justified on the following grounds:
first, by construction, $\mDp(K)$ is always positive and monotonically increasing with $K$;
second, it has the same asymptotic dependence on $K$ as $\mD$.
While a direct consequence of this decomposition is that $\Pip \to 0$ as $K\to \infty$,
this does not automatically guarantee that  {net flux}
{$\Pi(K)=\Pip(K) + \Pif(K)$}
is almost a
constant over a large range of scales at all $\De$.
Our numerical results thus provide an additional post-facto
justification of the decomposition of $\mP$. Also, we have checked that other reasonable choices
for $\mDp$ do not change the results qualitatively.}

For a moment, consider  again turbulence without polymers.
Assume that within the inertial range, in real space, the velocity shows scaling
behavior with an exponent $h$ such that, if we scale length by a factor of $b$,
$x\to bx$, then velocity scales as $u \to b^hu$.
In the inertial range, the flux equation, \eq{eq:eflux}, implies that the contribution
to the flux from the nonlinear term is constant.
Applying simple power-counting to the contribution to the flux from the nonlinear term,
we obtain $3h -1 = 0$, which implies the standard result from Kolmogorov theory
$h=1/3$.
Let us now apply the same scaling argument to the elastic range \cite{de2005homogeneous}.
As we scale $x\to bx$, we expect $u\to b^hu$, and $C \to b^g C$ with two distinct
exponents $h$ and $g$, respectively.
As the flux $\Pip$ is approximately constant in the elastic range, we obtain $h-1+g=0$.
By Fourier transform, it is straightforward to show that, if the velocity in real space
scales with an exponent $h$, then the scaling exponent for the energy,
$E(k)\sim k^{-\xi}$, with $\xi = 2h+1$.
Together the two relations imply that the scaling exponent for the shell-integrated
polymer energy is $\EP(k) \sim k^{-\psi}$, with $\psi = g +1 = 2-(\xi-1)/2 \approx 1.35$.
In \subfig{fig:spectra}{I} we plot the compensated shell-integrated polymer spectra from
our simulations; a scaling exponent of $\psi \approx 1.35$ is indeed consistent with our
results, independently corroborating the view of a polymer flux.

Next, we show how the flux-balance depends on the Deborah number in \subfig{fig:flux}{C-H}.
{We mark two Fourier modes in these plots, one is $\Kp$, marked by a black circle, the wavenumber at which $\Pif$ stops being the dominant contribution,
and the other is the wavenumber at which the dissipation ($\mD$ or $\mDp$) becomes the dominant term of the balance, which is a reasonable estimate
of $\keta$, marked by a black square.}
For small $\De$, \subfig{fig:flux}{C}, $\Kp > \keta$; in other words, the elastic range
is non-existent, masked by the viscous range.
As $\De$ increases, \subfig{fig:flux}{D-E},
$\Kp < \keta$ and the elastic range is clearly visible, with $\Kp$ reducing with $\De$.
As $\De$ increases beyond unity, $\Kp$ starts increasing again, \subfig{fig:flux}{F}, and
becomes almost equal to $\keta$ in \subfig{fig:flux}{G}.
For even larger $\De$, the elastic range disappears again. The values of $\Kp$ and $\keta$
obtained from \Fig{fig:flux} are used in \subfig{fig:spectra}{G--I} as the extension of the
scaling laws; the agreement between the two is an independent verification of the
validity of \eq{eq:pflux} and \eq{eq:pdiss}.

\begin{figure}
\centering
\includegraphics[width=0.45\textwidth]{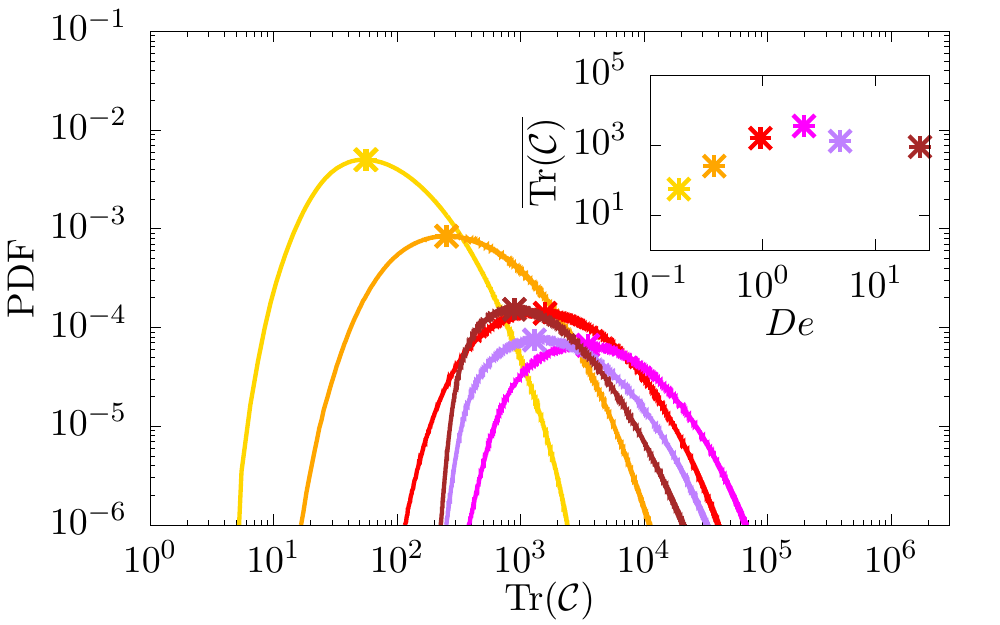}
\caption{\textbf{Polymer extension.}  Probability distribution function (PDF) of the {squared} extension  of the polymer for different Deborah numbers, measured in terms of $\Tr(\mC)$. The yellow, orange, red, magenta, purple and brown colors are used for increasing values of Deborah numbers, and {the color scheme is the same used in \Fig{fig:spectra}}. The mean polymer extension (marked with a cross) is a non-monotonic function of the Deborah number{, as shown in the inset of the figure}.}
\label{fig:pdf}
\end{figure}

Finally, this non-monotonic behavior of the scale $\Kp$ is also reflected in the
probability distribution function (PDF) of the {squared} extension of the polymers, $\Tr(\mC)$,
shown in \Fig{fig:pdf} for the Oldroyd-B model. 
For small Deborah numbers, the PDF has a peak somewhat higher than $3$, i.e., some polymers
are already not in a coiled state.
This is expected because the stretching of polymers is determined by the small scale
strain-rate~\cite{chertkov2000polymer,afonso2005nonlinear,musacchio2011deformation}
which is best captured by the Weissenberg number, which is about $16$ for the
smallest Deborah number we used. 
As $\De$ is increased, the peak of the PDF moves  to higher and higher values,
which is also what is expected.
Surprisingly, for  $\De > 1$ the peak moves back to smaller values.
This is an effect that cannot be captured from a passive polymer theory \cite{ben18} --
the feedback from the polymer to the flow changes the strain-rate such that
in turn the stretching of the polymers is reversed at Deborah number greater than
unity.
Note that here we show results from the Oldroyd-B model where
there is no constraint on the maximum stretching of polymers; however, this
non-monotonic behaviour is not unique to the Oldroyd-B model, and
we observe it also with the FENE-P model (see Fig.~(S4B) in the Supplementary Materials).
Furthermore, since the polymer extension does not continuously increase with the Deborah number,
the solution remains effectively dilute also at these high values of Weissenberg numbers, without
invalidating the dilute hypothesis of the models used.

To summarize, we have, so far, presented evidence from the largest resolution direct
numerical simulations of polymeric turbulence that, if the Deborah number lies in the
right range, $\Kp<k <\keta$, an elastic range with constant polymeric flux
$\Pip$ emerges in which $E(k) \sim k^{-\xi}$, with $\xi\approx 2.3$ and
$\EP(k) \sim k^{-\psi}$ with $\psi\approx 1.35$.
Crucially, the scale $\Kp$ behaves non-monotonically as a function of $\De$ and can
be precisely determined as the cross-over scale between $\Pif$ and $\Pip$.

\subsection{Structure function and intermittency}
\begin{figure*}
\centering
\includegraphics[width=0.95\textwidth]{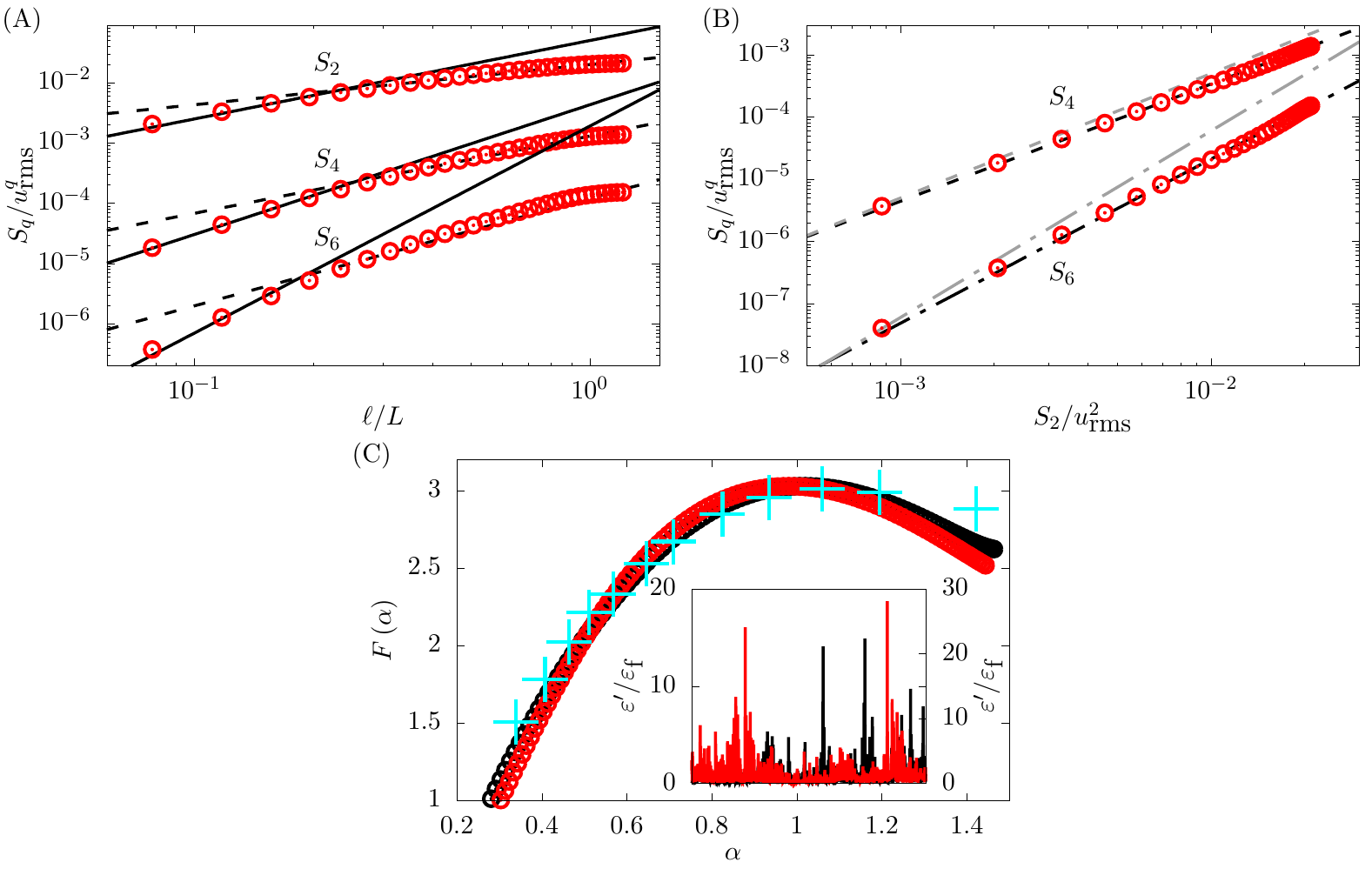}
\caption{\textbf{Intermittency.}
  (A): Structure function $\Sp$ for $q=2$, $4$ and $6$ from top to bottom for the viscoelastic case at $\De \approx 0.9$. The solid and dashed lines represent the expected scalings in the polymer and inertial range of scales, corrected by the intermittency correction.
  (B): Same structure functions of the panel (A), plotted in their
  Extended Self Similarity form {\cite{benzi_1993aa}}. The black and gray lines represent the expected scaling
  with and without intermittency corrections. The polymer and inertial range of scales
  follows the same line, indicating that the intermittency in the two ranges is the same.
  (C): Multifractal spectrum of the fluid dissipation. The cyan symbols represents
  experimental data taken from Ref.~\cite{meneveau1991multifractal}.
  The inset shows typical signals for the fluid dissipation,
  normalized by its mean value.
  In the two figures, the Newtonian and viscoelastic cases are shown in black and red,
  respectively.
}
\label{fig:intermittency}
\end{figure*}

In the absence of polymers, the scaling exponents of the structure function $\zetaq$ are a
nonlinear function of $q$ -- a phenomena known as intermittency~\cite[][Chapter 8]{Fri96},
which can be parametrised by corrections to the Kolmogorov scaling
\begin{equation}
  \zetaq = q/3 + \deltaq \/.
  \label{eq:zetaq}
\end{equation}
The best estimates~\cite{sreenivasan1998there} of $\deltaq$ are $\delta_{\rm 2} \approx 0.04$,
$\delta_{\rm 4}\approx -0.05$, $\delta_6 \approx -0.23$, whereas
$\delta_{\rm 3}  = 0$ due to Kolmogorov's
four-fifth law.
We now explore what happens to intermittency {\it on the addition of polymers.}
In \subfig{fig:intermittency}{A} we plot the structure function for $q=2,4,$ and $6$
for $\De \approx 0.9$ -- the case for which we have the largest elastic range.
The second order structure function, \eq{eq:zetap}, with $q=2$, is the Fourier transform
of the energy spectrum $E(k)$.
Hence, if $E(k) \sim k^{-\xi}$, then $\Stwo(\ell) \sim \ell^{\zetat}$, with
$\zetat = \xi-1 \approx 1.3$, which is what we obtain.
On the other hand, the scalings for $q=4$ and $q=6$ are different from
$2\zetat$ and $3\zetat$;
this becomes obvious when we plot in \subfig{fig:intermittency}{B} $S_4$ and $S_6$ as a
function of $\Stwo${\cite{benzi_1993aa}}.
In these plots the elastic range and the inertial range seems to merge into one
scaling range, suggesting that the intermittency correction for $q=4$ and $q=6$ are
the same in both the elastic and the inertial range.
Our results on intermittency discussed so far, agree with the
experimental results obtained in Ref.~\cite{zhang2021experimental}.
{Thus we must conclude that 
the effect of the polymers is to change the dominant exponent $q/3$ but
not the intermittency correction $\delta_q$ !  The dominant exponent is
determined by the scaling of the mean value of the energy flux, $\Pif$,
whereas the intermittency exponents are determined by
the fluctuations of the energy flux~\cite[][Chapter 8]{Fri96}.
The polymers change the mean significantly, but
the fluctuations are still dominated by the fluctuations of viscous energy
dissipation which remains unchanged on the addition of polymers.}

{To check this hypothesis we now use
 an alternative way to explore intermittency: through the statistics
 of the viscous dissipation.
 We find that the $q$-th moment of the viscous dissipation averaged over a ball of radius $\ell$ shows scaling, viz., 
\begin{subequations}
  \begin{align}
  \bra{\epsl^q} &\sim \ell^{\lamq}, \quad\text{where}\\ 
  \epsl &\equiv \frac{2\muf}{\rhof}\bra{\Sab\Sab}_{\ell}\/.
  \label{eq:epsfl}
  \end{align}
\end{subequations}
Here the symbol $\bra{\cdot}_{\ell}$ denotes averaging over a ball
of radius $\ell$ and the symbol $\bra{\cdot}$ averaging over the
statistically stationary state of turbulence.
For $\ell=L$,  $\bra{\epsilon_L}=\epsilonf$ the viscous dissipation in\eq{eq:eps}.
The Legendre transform of the function $\lamq$ give the multifractal
spectrum of turbulence $F(\alpha)$ which we plot in \subfig{fig:intermittency}{C}.
Our results, for the Newtonian case, agrees with the experiments in the Newtonian turbulence  \cite{meneveau1991multifractal}.
Remarkably, we find that the multifractal spectrum is the same with or without
polymers, thereby confirming our hypothesis. }

{Altogether these evidences point towards the scenario. 
For small enough viscosity and for small enough $k$ ($k_{inj}< k <k_d$), the energy flux has two
contributions -- the advective flux and the polymeric flux.
  In the inertial range the advective flux dominates. In the elastic range
  the polymeric flux dominates. However the intermittency exponents are determined
  not by the mean value of the flux but by its fluctuations. The fluctuations
  are determined by the fluctuations of the viscous dissipation, which
  remains the same in both polymeric and Newtonian turbulence. }

\section{Discussion}
Our simulations reach the highest $\Rey$ and $\De$ numbers reached so far in
numerical simulations of homegeneous and isotropic turbulence of polymer
solutions~\footnote{See supplemental material for a
  comparisons of our parameters with all the earlier simulations.}.
Hence, by modern nomenclature, we may call this
\textit{elasto-inertial turbulence}~\cite{samanta2013elasto,valente2016energy},
which is merely a renaming of the traditional field of polymeric turbulence.
We find that the central role of the polymers is that the cascade of energy,
which in absence of polymers is determined by the advective nonlinearity,
is now carried by both the advective nonlinearity and the polymer stress
\textit{but at different scales}.
At large scales, the energy flux through scales is dominated by the advective
nonlinearity, while the polymer stress plays a sub-dominant role --
this is reversed at smaller scales.
This gives rise to two scaling ranges, the classical Kolmogorov one and the
new elastic one. 
We emphasize that the new scaling we find is a purely elastic effect,
the advective nonlinearity plays a sub-dominant role in the range of
scales where the elastic scaling is observed. {A comparison of different models 
of polymeric fluids confirms that elasticity and not shear thinning is crucial to 
observe the new elastic scaling range in the energy spectrum.}
Thus elasto-inertial turbulence appears to be inertial turbulence at large
scale and a new elastic behaviour -- different from elastic turbulence --
at smaller scales which are still larger than viscous scales.
The viscous effects may dominate over the elastic effects for small $\Rey$,
thereby making the elastic range disappear.
Furthermore, we establish that this elastic behaviour is non-monotonic
in Deborah number. {A simple qualitative explanation for this effect is that when $\De \gg 1$,
polymers are not able to properly stretch due to their timescale being much larger than the
largest timescale of the fluid, thus acting as a filter of the velocity fluctuations.  
However, our simulations with passive polymers show that this scenario is not true (polymer extension 
increases monotonically with $\De$). Thus the non-monotonic behavior observed by us cannot be captured 
by a theory that treats polymers as passive objects.}


At present there are no theories that help us understand the novel scaling.
To the best of our knowledge the first theory that predicted a new power-law
scaling in elastic range is by Bhattacharjee and
Thirumalai~\cite{bhattacharjee1991drag,thirumalai1996polymer},
whose theory gives an exponent of $\xi = 3$ in the elastic range;
by contrast we observe $\xi \approx 2.3$, consistent with recent
experiments~\cite{zhang2021experimental}.
Bhattacharjee and Thirumalai also assumed that most of the polymers
have not undergone coil-stretch transition.
In our simulations, this may be true at small $\De$, where the elastic range
is non-existent, but this is definitely not the case at high $\De$ where
we do observe the elastic scaling. 
The theory by Fouxon and Lebedev~\cite{fux03} also predicts a power-law scaling
($\xi > 3$) and the existence of an elastic range, but we agree with
Zhang et al~\cite{zhang2021experimental} that
``the assumptions and quantitative prediction of the theory are not supported by''
our numerics.
{We believe the  elastic range we observe is distinct from elastic turbulence
in two ways. One, the scale--dependent Reynolds number in the elastic range
is not necessarily very small.
Two, we find $\xi \approx 2.3$ whereas 
almost all study of elastic turbulence find $\xi > 3$~\cite{gro+ste00,
  steinberg2021elastic, berti2006small, watanabe2014power,ray2016elastic,
  gupta2017melting},
consistent with the theory of Fouxon and Lebedev~\cite{fux03}.
Note that,  at least one other simulation of elastic
turbulence~\cite{gupta2019effect} has found $\xi < 3$ 
in two dimensional polymeric flows.  
}

Our simulations extends the recent experiments by
Zhang et. al.~\cite{zhang2021experimental}, who did not probe the
Deborah number dependence, by
measuring quantities that are not easily accessible in the experiments,
e.g., the contribution from the polymeric stress and the PDF of polymer extension,
thereby providing constraints and clues to a future theory.
We show, for the first time, that the polymer contribution can be decomposed into a
purely dissipative term and into a purely energy flux, with
the latter transporting the majority of energy in the elastic range.
Its validity has been confirmed in several ways:
\textit{(i)} its span is consistent with the range of the elastic scale in the
energy spectra;
\textit{ii)} the polymer energy spectra exhibits a scaling consistent in range
and slope with it.
{Finally, we show that the intermittency corrections are the same in the elastic and
the Newtonian cases. This indicates that the statistical nature of the fluctuations
of the energy flux remains unchanged on addition of polymers -- the fluctuations
  are determined by the statistics  of the viscous energy dissipation, which
  remains the same in both polymeric and Newtonian turbulence.}

\section{Materials and methods}

The viscoelastic fluid is governed by the conservation of momentum and the incompressibility constraint:
\begin{subequations}
\begin{align}
\label{eq:NS}
\rhof \left(\frac{\partial \ua}{\partial t}
+ \frac{\partial \ua\ub}{\partial \xb} \right) &=
- \frac{\partial p}{\partial \xa} +
\frac{\partial}{\partial \xb}\left( 2 \muf S_{\ab} 
+  \frac{\mup}{\taup} f \Cab\right)\/, \\
\frac{\partial \ua}{\partial \xa} &= 0\/.
\end{align}
\end{subequations}
In the previous set of equations, $\rhof$ and $\muf$ are the density and dynamic viscosity of the fluid, $p$ is the pressure, and $\mS$ the rate-of-strain tensor with components $S_{\ab}$ defined as $S_{\ab}=\left( \partial \ua/ \partial \xb + \partial \ub / \partial \xa\right)/2$. The last term in the momentum equation is the non-Newtonian contribution, with $\mup$ being the polymer viscosity, $\taup$ the polymer relaxation time, $f$ a scalar function and $\mC$ the conformation tensor with components $\Cab$ found by solving the following transport equation:
\begin{equation}
\label{eq:adv}
\frac{\partial \Cab}{\partial{t}} + u_\gamma \frac{\partial C_{\alpha \beta}}{\partial x_\gamma} =
 C_{\alpha\gamma}\delg{\ub} +C_{\gamma\beta}\dela{\ug} -\frac{f\Cab - \delta_{\ab}}{\taup}\/.
\end{equation}
The function $f$ is equalt to $f=1$ in the purely elastic Oldroyd-B model, and to $f = \left(\mL^2-3 \right)/\left(\mL^2-C_{\gamma\gamma}\right)$ in the FENE-P model ($\mL$ is the maximum polymer extensibility) exhibiting both shear-thinning and elasticity. Turbulence is sustained by an additional forcing in the momentum equation; in particular, we use the spectral scheme by~\citet{eswaran1988forcing} to randomly injecting energy within a low-wavenumber shell with $1\le k \le2$.

The equations of motion are solved numerically within a periodic cubic domain box of length $2\pi$, discretized with $\tN=1024$ grid points per side with a uniform spacing in all directions, resulting in a total number of around $1$ billion grid points. The grid resolution $\kmax$ used in the present work is the largest used for viscoelastic fluids and is sufficient to represent all the relevant quantities of interest till the Kolmogorov length-scale $\eta$ ($\kmax\eta \approx 1.7$) {\cite{pope_2001a}. Furthermore, the smallest temporal scale of the flow, i.e. the Kolmogorv time-scale $\tau_\eta$, is overly resolved (by two order of magnitude $\tau_\eta/\Delta t \approx 600$), due to stability constraint arising from the non-Newtonian features of the flow, strongly increasing the computational cost.} {We have confirmed that these results are robust with respect to change in spatial and temporal resolutions,
as reported in Fig.~(S2)
where the energy spectra obtained by different time and space resolutions are compared, finding the robustness of the reported results}. To solve the problem, we use the flow solver \textit{Fujin}, an in-house code, extensively validated and used in a variety of problems~\cite{rosti2019droplets,rosti2020increase,rosti2020fluid,olivieri2020dispersed,rosti2021turbulence,mazzino_rosti_2021b,brizzolara2021fibre}, {based on the (second-order) finite-difference method for the spatial discretization and the (second-order) Adams-Bashforth scheme for time marching.} See also  \texttt{https://groups.oist.jp/cffu/code} for a list of validations. {The non-Newtonian stress equation is solved following the (exact) log-conformation approach \cite{fattal_kupferman_2004a} to ensure the positive-definiteness of the tensor even at high $\De$, without the addition of any artificial stabilising terms.}

\section{References and notes}
\bibliography{turb_ref}

\section{Acknowledgments}
M.E.R. thanks Ms. Megumi Ikeda of the Complex Fluids and Flows unit at OIST for the
help and useful discussions in preparing the flow visualisation.

\subsection{Funding}
M.E.R. is supported by the Okinawa Institute of Science and Technology Graduate University (OIST) with subsidy funding from the Cabinet Office, Government of Japan. M.E.R. also acknowledges the computational time provided by HPCI on the Fugaku cluster under the grants hp210229 and hp220099, and the computer time provided by the Scientific Computing section of Research Support Division at OIST.
PP  acknowledges support from the Department of Atomic Energy (DAE), India under Project Identification No. RTI 4007, and DST (India) Project Nos. ECR/2018/001135 and DST/NSM/R\&D\_HPC\_Applications/2021/29.
DM acknowledges the support of the Swedish Research Council Grant No. 638-2013-9243 and 2016-05225.

\subsection{Author contributions}
M.E.R. conceived the original idea, planned the research, developed the code and
performed the numerical simulations. All authors analyzed data, outlined the
manuscript content and wrote the manuscript.
\subsection{Competing interests}
The authors declare that they have no competing interests.
\subsection{Data availability}
All data needed to evaluate the conclusions are present in the paper and/or the Supplementary Materials. 

The code used for the present research is a standard direct numerical simulation solver for the Navier--Stokes equations. Full details of the code used for the numerical simulations are provided in the Methods section and references therein.  

\newpage

\section*{Supplementary information}

\setcounter{table}{0}
\makeatletter 
\renewcommand{\thetable}{S\@arabic\c@table}
\makeatother

\setcounter{figure}{0}
\makeatletter 
\renewcommand{\thefigure}{S\@arabic\c@figure}
\makeatother

\subsection*{Energy balance equation}
\label{app:spectral-balance}
{We perform the Fourier transform of the governing equations to obtain an expression for the turbulent kinetic energy spectrum $\hat{E}(\boldsymbol{\kappa},t) \equiv \frac{1}{2}\rho\langle \boldsymbol{\hat{u}}^{*} \cdot \boldsymbol{\hat{u}}\rangle$, where $(\hat{.})$ denotes the Fourier transform into the spectral space, $\boldsymbol{\kappa}$ denotes the wave vector with a magnitude $\kappa$, and the superscript $*$ denotes the complex conjugate;
\begin{equation}\label{eq:massF}
  \boldsymbol{\kappa} \cdot \boldsymbol{\hat{u}}=0,
\end{equation}
\begin{equation}\label{eq:momF}
	\rho\frac{\mathrm{d} \boldsymbol{\hat{u}}}{\mathrm{d} t} + \boldsymbol{\hat{G}} = -i \boldsymbol{\kappa} \hat{p} - \mu_\text{f}{\kappa^{2}}\boldsymbol{\hat{u}} + \boldsymbol{\hat{F}}_\text{pol} + \boldsymbol{\hat{F}},
\end{equation}
where $\hat G$ is the Fourier coefficient of the non-linear convective term appearing in the momentum equation, and $i$ is the imaginary unit. Similar equations can be obtained for the complex conjugate $\hat {\boldsymbol{u}}^{*}$. When the momentum equation is multiplied by $\boldsymbol{\hat{u}}^{*}$, the pressure term $-i\boldsymbol{\kappa} \cdot \boldsymbol{\hat{u}}^{*} \hat{p}$ vanishes due to the incompressibility constraint, and the viscous term $- \mu_\text{f}{\kappa^{2}}\boldsymbol{\hat{u}}\cdot \boldsymbol{\hat{u}}^{*}$ can be expressed in terms of the kinetic energy; $-2 \mu_\text{f}{\kappa^{2}} \hat{E}$. The same holds when multiplying the momentum equation of $\hat {\boldsymbol{u}}^{*}$ by $\hat {\boldsymbol{u}}$. By summing the two equations for $\hat {\boldsymbol{u}}$ and $\hat {\boldsymbol{u}}^{*}$ and dividing by $2$, we have an expression for the time evolution of turbulent kinetic energy $\hat{E}(\boldsymbol{\kappa},t)$
\begin{equation}\label{eq:Espec}
	\frac{\mathrm{d} \hat{E}(\boldsymbol{\kappa})}{\mathrm{d} t} =  \hat{T}(\boldsymbol{\kappa}) + \hat{V}(\boldsymbol{\kappa}) + \hat{F}_{\text{pol}}(\boldsymbol{\kappa}) + \hat{F}(\boldsymbol{\kappa}) ,
\end{equation}
where the terms on the right-hand side represent the following contributions: $\hat{T}= -\frac{1}{2}(\boldsymbol{\hat{G}}\cdot \boldsymbol{\hat{u}}^{*}+\boldsymbol{\hat{G}}^{*}\cdot \boldsymbol{\hat{u}})$ is due to the non-linear convective term, $\hat{V} = - 2 \mu_\text{f}{\kappa^{2}} \hat{E}$ is due to the fluid dissipation term, $\hat{F}_{\text{pol}}=\frac{1}{2}(\boldsymbol{\hat{F}_{\text{pol}}}\cdot \boldsymbol{\hat{u}}^{*}+\boldsymbol{\hat{F}_{\text{pol}}^{*}}\cdot \boldsymbol{\hat{u}})$ is due to the  non-Newtonian stress, and $\hat{F}_\text{inj}=\frac{1}{2}(\boldsymbol{\hat{F}}_\text{inj}\cdot \boldsymbol{\hat{u}}^{*}+\boldsymbol{\hat{F}}_\text{inj}^{*}\cdot \boldsymbol{\hat{u}})$ is due to the external forcing. The one-dimensional energy spectrum $E(\kappa,t)$ can be obtained by isotropically averaging~\eqref{eq:Espec} over the sphere of radius $\kappa$ (i.e., $E(\kappa,t)=\iint_{S(\kappa)} \hat{E}(\boldsymbol{\kappa},t) \mathrm{d}S(\kappa)$, where $S(\kappa) $ is the sphere defined by $\boldsymbol\kappa\cdot\boldsymbol\kappa= \kappa^{2}$),
\begin{equation}\label{eq:Espec_avg}
	\frac{\mathrm{d} {E(\kappa)}}{\mathrm{d} t} =  T(\kappa) + V(\kappa) + F_{\text{pol}}(\kappa) + F_\text{inj}(\kappa).
\end{equation}
where the time derivative becomes zero for a statistically stationary flow. Integrating the equation above from $\kappa$ to infinity, we obtain the energy-transfer balance
\begin{equation}
	\label{eq:E_balance}
	0 = \Pif + \mD' + \mathcal{P}' + \mathcal{F}_\text{inj},
\end{equation}
where $\Pif(\kappa) \equiv \int_\kappa^\infty T (\kappa) \, \mathrm{d}\kappa$,\,\, $\mD'(\kappa) \equiv \int_\kappa^\infty V(\kappa) \, \mathrm{d}\kappa$,\,\,  $\mathcal{F}_\text{inj}(\kappa) \equiv \int_\kappa^\infty F_\text{inj}(\kappa) \, \mathrm{d}\kappa$, and $\mathcal{P}'(\kappa) \equiv \int_\kappa^\infty F_\mathrm{pol}(\kappa) \, \mathrm{d}\kappa$ represent the contributions to the spectral power balance from the non-linear convective, fluid dissipation, turbulence forcing, and non-Newtonian terms, respectively. The fluid dissipation term can be expressed as $\mD(\kappa) = -\int_0^\kappa V(\kappa) \, \mathrm{d}\kappa=\mD'(\kappa)+{\langle\epsilonf\rangle}$, where ${\langle\epsilonf\rangle}=-\int_0^\infty V(\kappa) \, \mathrm{d}\kappa$ is the rate of energy dissipated by the fluid viscosity. Similarly, the non-Newtonian contribution can be written as $\mathcal{P}(\kappa) = -\int_0^\kappa F_{\text{pol}}(\kappa) \, \mathrm{d}\kappa=\mathcal{P}'(\kappa) + {\langle\epsilonp\rangle}$, where ${\langle\epsilonp\rangle}=-\int_0^\infty F_{\text{pol}}(\kappa)  \, \mathrm{d}\kappa$ is the non-Newtonian dissipation rate. Substituting these in the above equation, we obtain the energy balance equation used in the main document.}

\begin{figure}
\centering
\includegraphics[width=0.45\textwidth]{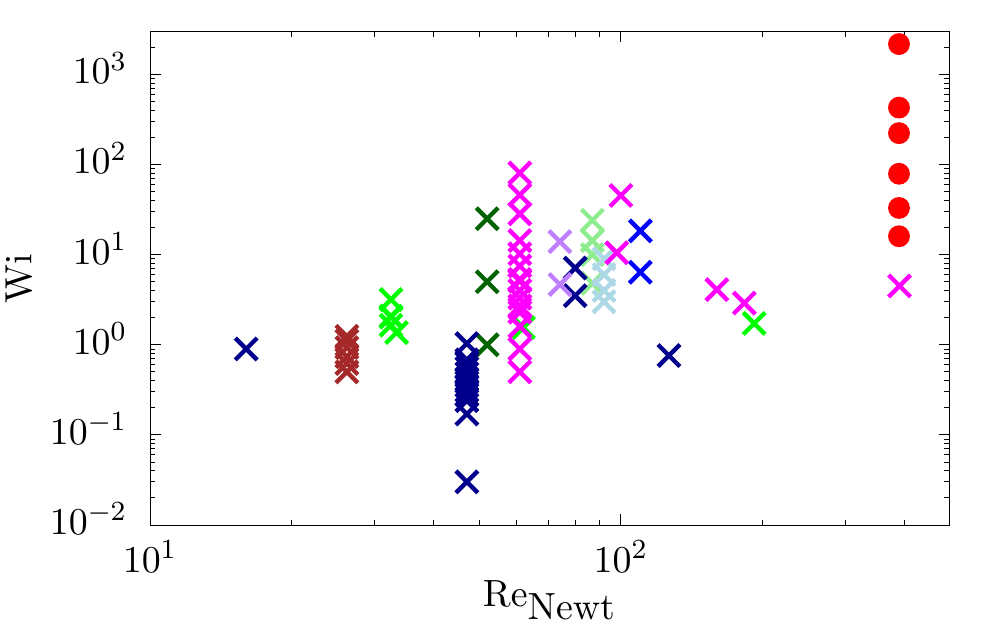}
\caption{\textbf{DNS data of polymeric turbulence available in the literature.}  The plot reports the reference Reynolds numbers of the Newtonian cases $\Rey_{\mbox{Newt}}$ and the Weissenberg numbers $\Wi$ of the simulations available in the literature: the cross symbols represents the previous investigations, while the filled circles the present one. Light green: \citet{berti2006small}; green: \citet{nguyen2016small}; dark-green: \citet{watanabe2013hybrid,watanabe2014power}; light blue: \citet{fathali2019spectral}; blue: \citet{de2012control}; dark blue: \citet{per+mit+pan06,per+mit+pan10}; magenta: \citet{valente2014effect,valente2016energy}; purple: \citet{de2005homogeneous}; brown: \citet{cai2010dns}. The results of the present work focus on a high $\Rey$ and $\Wi$ region never investigated before.}
\label{fig:DNSdata}
\end{figure}

\begin{figure}
\centering
\includegraphics[width=0.45\textwidth]{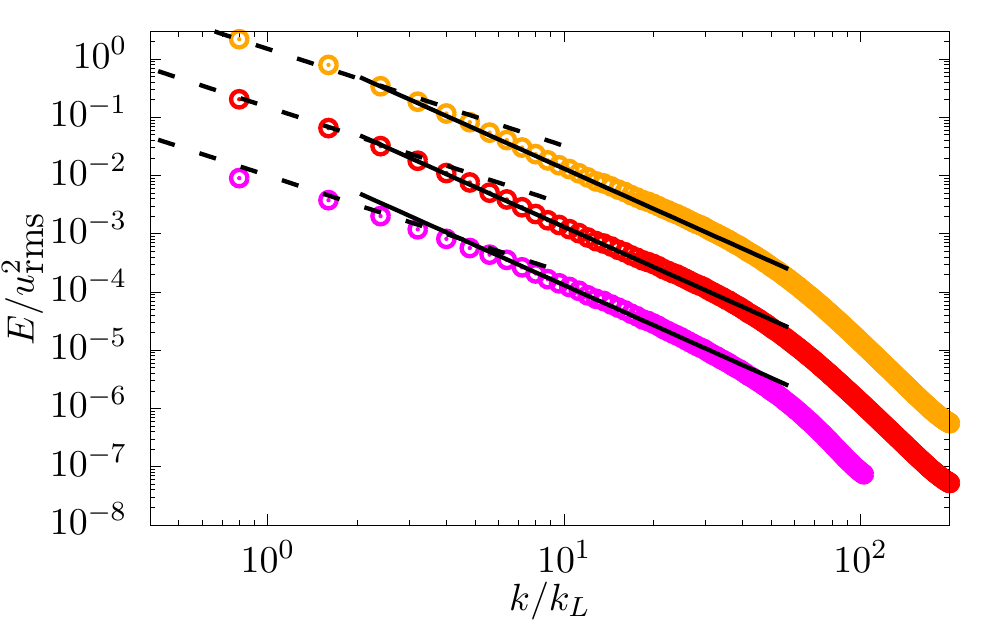}
\caption{{\textbf{Numerical resolution.}  Energy spectra for $De \approx 0.95$ obtained by (orange) reducing the timestep by a factor $10$ and by (magenta) reducing the grid size by a factor $2$. The spectra are shifted vertically for visual clarity. In both cases, the grid and time resolution proves to be appropriate.}}
\label{fig:dxt}
\end{figure}

\subsection*{Total energy budget}
{The total energy of a FENE-P polymeric fluid system described by the governing 
equations is given by~\cite{de2005homogeneous}:
\begin{equation}
	F \equiv \underbrace{ \frac{1}{2}\bra{\lvert\uu\rvert^2}}_{\mEf}
	+ \underbrace{ \frac{\mup}{2\rhof \taup}
	          \left[\bra{(L^2-3)\log(f)} + 3\right] }_{\mEp},
\label{eq:entot}
\end{equation}
where $\mEf$ is the kinetic energy, and $\mEp$ is the contribution due to the polymer 
conformation. 
Note that by taking the limit $L\to \infty$ we get the corresponding expression for 
the Oldroyd-B model, 
$\mEp=(\mup/2\rhof\taup) \bra{ C_{\gamma\gamma}} $. 
By taking the time-derivative of \eqref{eq:entot}, and using equations \eqref{eq:ns}, 
\eqref{eq:poly} we get the following energy budget equation
\begin{equation}
\frac{\partial F}{\partial t} = 
	-\underbrace{\frac{2 \muf}{\rho_f} \langle S_{\alpha \beta} S_{\alpha \beta} \rangle}_{\epsilonf}  
	- \underbrace{\frac{\mup}{2 \rhof\taup^2} \bra{ f (f C_{\mu \mu}-3)}}_{\epsilonp}.
\end{equation}}

\subsection*{Elasticity and shear-thinning}
\begin{figure}
\centering
\includegraphics[width=0.95\textwidth]{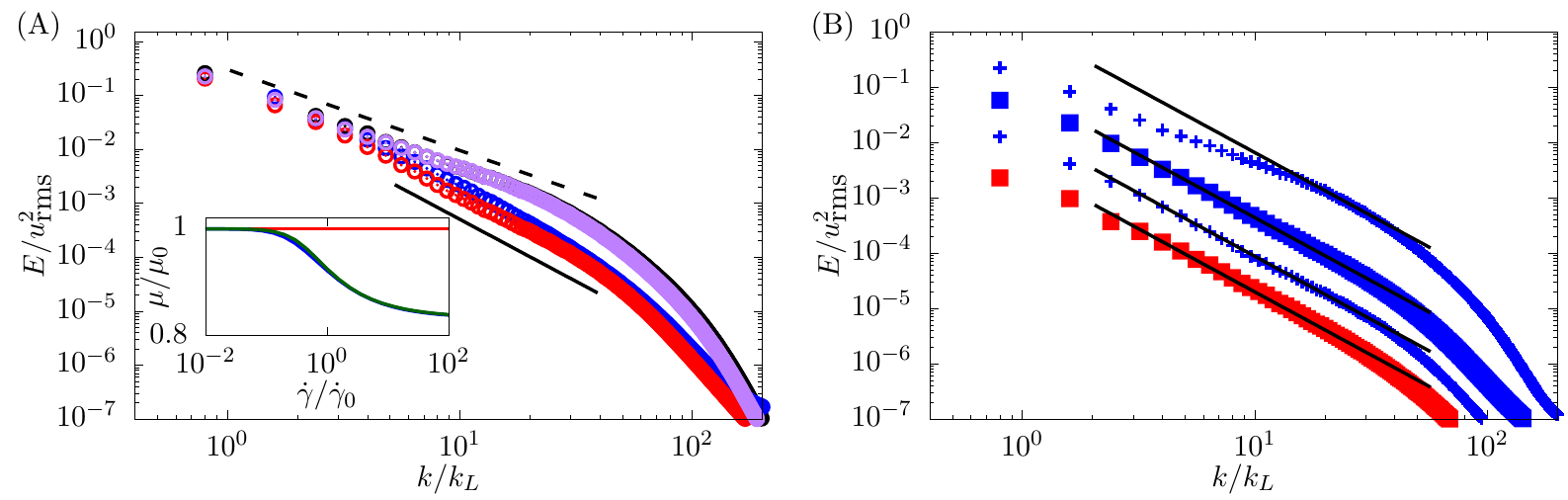}
\caption{\textbf{Effect of shear-thinning and elasticity on the energy spectra.}
  (A): Energy spectra for $\De \approx 0.95$ for different fluids.
  Black: Newtonian fluid. Purple: inelastic Carreau Yasuda model.
  Blue: FENE-P model. Red: Oldroyd-B model.
  The inset  shows the shear rheology of the same fluids.
  The new scaling at intermediate small scales is a purely elastic effect,
  which completely disappear in the absence of elasticity while it is reduced
  when shear-thinning is present together with elasticity.
  {(B): Energy spectra for $\De \approx 0.95$ for FENE-P fluids with different
    $\mL$.
 From top to bottom: $\mL= 20$, $60$, $100$, $\infty$ (Oldroyd-B model).  The spectra in the right panel are shifted vertically for visual clarity.
  The new scaling at intermediate small scales is present for all cases with sufficiently large $\mL$.}
  The dashed and solid lines represent the scaling $k^{-\psi}$ with $\psi=5/3$ and $\psi\approx 1.35$.
}
\label{fig:spectraST}
\end{figure}

We consider three different models of polymeric fluids: the Oldroyd-B model (elasticity), the FENE-P model (elasticity and shear-thinning), and the Carreau--Yasuda model (shear-thinning), see Tab.~\ref{tab:param} for the full list of simulations performed. In the inealstic shear-thinning fluid, the fluid viscosity $\muf$ is a function of the local shear rate $\dot{\gamma}$ as
\begin{equation}
\frac{\mu}{\mu_0} = \frac{\mu_\infty}{\mu_0} + \left( 1 - \frac{\mu_\infty}{\mu_0} \right) \left[ 1+ \left( \taup \dot{\gamma} \right)^2 \right]^{\left( \frac{n-1}{2} \right)}\/,
\end{equation}
where $\mu_0$ and $\mu_\infty$ are the viscosity at zero and infinite shear rates, $n$ is the the power index ($n=0.4$) and $\taup$ the consistency index. The model parameters are found to fit the FENE-P shear-thinning rheology, as shown in the inset of \subfig{fig:spectraST}{A}.

\begin{figure}
\centering
\includegraphics[width=0.95\textwidth]{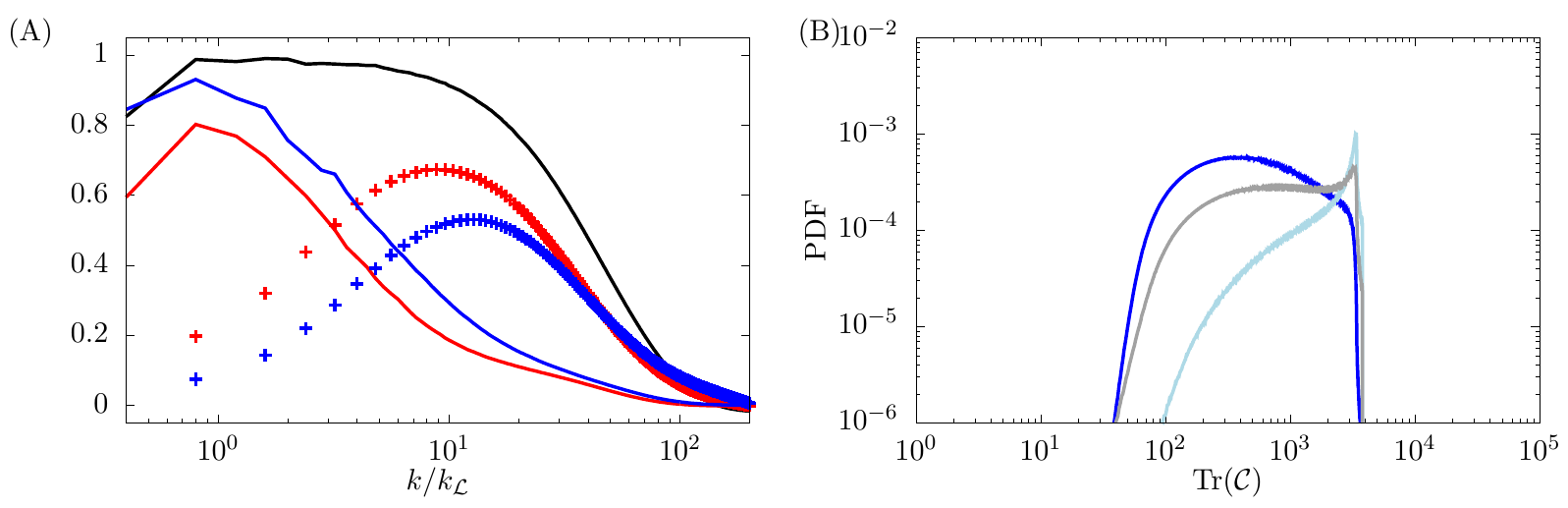}
\caption{\textbf{Effect of shear-thinning on the energy balance and polymer extension.}  
  {(A): the non-linear energy flux $\Pif$ (solid line) and polymer flux, $\Pip$ ($+$) for the (black) Newtonian, (red) Oldroyd-B and (blue) FENE-P fluids at $De \approx 1$. In the FENE-P fluid, the non-linear energy flux is stronger than in the Oldroyd-B case, resulting in a reduced polymer flux.} (B): Probability distribution function (PDF) of the polymer extension for different Deborah numbers, measured in terms of $\Tr(\mC)$, obtained with the FENE-P model. Blue: $\De \approx 0.95$; light blue: $\De \approx 2.31$; gray: $\De \approx 4.82$. Differently from the results obtained with the Oldroyd-B model, the polymer extension is upper bounded by $\mL^2$; however, the non-monotonic trend with $\De$ is still evident.
   }
\label{fig:pdfST}
\end{figure}

{When comparing the results obtained for different non-Newtonian models, we find that the new power-law scaling in the energy spectra at intermediate small scales is a purely elastic effect, which completely disappear in the absence of elasticity while it is slightly reduced in range when shear-thinning is present together with elasticity, as shown in \subfig{fig:spectraST}{A}. When both elasticity and shear-thinning effects are present, the exponent of the power-law scaling remains practically unchanged if the parameter $\mL$ (the maximum possible extension of the polymers) of the FENE-P model is varied within a reasonable range, while the elastic scaling range can completely disappear for too small values, showing a complex behaviour. While the presence of shear-thinning in combination with elasticity does not alter the slope of the elastic range, it reduces its range: this is caused by an enhancement of the non-linear energy flux $\Pif$ and a consequent reduction of the polymer flux, $\Pip$, as shown in \subfig{fig:pdfST}{A}.}
{In the FENE-P model the extension of the polymers is arrested due to two mechanisms.
  One because of the feedback to the flow, and two because of the nonlinear saturation
  term in the FENE-P equation itself. If $\mL$ is small, the nonlinear saturation
  term stops the polymers from having large extensions. This implies that the
  feedback from the polymer to the flow is also small, consequently the flux $\Pip$
  is also small. Hence, as $\mL$ is increased, $\Pip$ increases and $\Pif$ decreases.
  When $\mL \to \infty$, which corresponds to the Oldroyd-B model, we get the largest
  possible values for $\Pip$. In this case the elastic range is best observed. }
\begin{table*}[h]
\caption{{List of all cases considered in the present work.}}
\label{tab:param}
\begin{tabular}{cccccccc}
\hline
case & model & $\Rel$ & $\De$ & $\Wi$ & $\mup/\left(\mup+\muf \right)$ & $\mL$\\
\hline
$\ocircle$ & single phase & $390$ & $-$ & $-$ & $0.1$ & $-$\\
\hline
\textcolor{yellow}{$\pentagon$}  & Oldroyd-B & $480$ & $0.18$ & $16$   & $0.1$ & $\infty$\\
\textcolor{orange}{$\lozenge$}  & Oldroyd-B & $630$ & $0.37$ & $33$   & $0.1$ & $\infty$\\
\textcolor{red}{$\ocircle$}     & Oldroyd-B & $740$ & $0.95$ & $79$   & $0.1$ & $\infty$\\
\textcolor{magenta}{$\vartriangle$} & Oldroyd-B & $690$ & $2.31$ & $223$  & $0.1$ & $\infty$\\
\textcolor{violet}{$\triangledown$}  & Oldroyd-B & $614$ & $4.82$ & $427$  & $0.1$ & $\infty$\\
\textcolor{brown}{$\Square$}   & Oldroyd-B & $447$ & $24.6$ & $2169$ & $0.1$ & $\infty$\\
\hline
\textcolor{blue}{$+$}      & FENE-P    & $420$ & $0.95$ & $79$   & $0.1$ & $20$\\
\textcolor{blue}{$\ocircle$}      & FENE-P    & $610$ & $0.95$ & $79$   & $0.1$ & $60$\\
\textcolor{blue}{$\times$}      & FENE-P    & $700$ & $0.95$ & $79$   & $0.1$ & $100$\\
\textcolor{Aquamarine}{$\ocircle$}      & FENE-P    & $620$ & $2.31$ & $31$   & $0.1$ & $60$\\
\textcolor{gray}{$\ocircle$}      & FENE-P    & $487$ & $4.82$ & $31$   & $0.1$ & $60$\\
\hline
\textcolor{purple}{$\ocircle$} & Carreau-Yasuda & $370$ & $-$ & $-$ & $-$ & $-$\\
\hline
\end{tabular}
\end{table*}

\end{document}